\documentclass[12pt]{article}
\usepackage{latexsym}
\newcommand{\be}{\begin{equation}}
\newcommand{\ee}{\end{equation}}

\newcommand{\bea}{\begin{eqnarray}}
\newcommand{\eea}{\end{eqnarray}}

\newcommand{\we}{\wedge}

\newcommand{\p}{\partial}
\def\id{\protect{{1 \kern-.28em {\rm l}}}}

\def\we{{\wedge}}

\input amssym

\def\oneone{\rlap 1\mkern4mu{\rm l}}
\def\coeff#1#2{\relax{\textstyle {#1 \over #2}}\displaystyle}

\def\IR{\Bbb{R}}
\begin{document}

 \begin{titlepage}

\begin{flushright}
UCLA/04/TEP/31\\
hep-th/0408106
\end{flushright}

\bigskip 
\bigskip
\centerline{\Large \bf One Ring to Rule Them All ...}
\medskip
\centerline{\Large \bf and in the Darkness Bind Them? }
\bigskip
\bigskip
\centerline{{\bf Iosif Bena$^1$ and Nicholas P. Warner $^2$}}
\medskip
\centerline{$^1$ Department of Physics and Astronomy}
\centerline{University of California}
\centerline{Los Angeles, CA  90095, USA }
\medskip
\centerline{$^2$ Department of Physics and Astronomy}
\centerline{University of Southern California}
\centerline{Los Angeles, CA 90089, USA}
\bigskip
\centerline{{\rm iosif@physics.ucla.edu,~~ warner@usc.edu} }
\bigskip \bigskip

\begin{abstract}

We construct all eleven-dimensional, three-charge BPS solutions that
preserve a fixed, standard set of supersymmetries.  Our solutions
include all BPS three-charge rotating black holes, black rings,
supertubes, as well as arbitrary superpositions of these objects. We
find very large families of black rings and supertubes with profiles
that follow {\it arbitrary} closed curves in the spatial $\IR^4$
transverse to the branes. The black rings copiously violate black hole
uniqueness. The supertube solutions are completely regular, and
generically have small curvature. They also have the same asymptotics
as the three-charge black hole; and so they might be mapped to
microstates of the D1-D5-p system and used to explain the entropy of
this black hole.

\end{abstract}

\end{titlepage}

\section{Introduction}

One of the more surprising features of supergravity is the existence
of large families of supersymmetric solutions that preserve the same 
class of supersymmetries. Some of these families of solutions arise
from D-brane configurations that are assembled so as to create supertubes
and black holes.  The supersymmetries of these solutions are then 
determined by the canonical projectors associated with each set of D-branes.

For the D1-D5 system, the smooth bulk supergravity solutions dual to 
microstates of the boundary CFT have  been classified, and 
shown to precisely account for the entropy of this theory \cite{mathur}.
While these results are extremely interesting
and suggestive descriptions of black hole microstates, the D1-D5
system does not give rise to ``true'' black holes with non-zero
horizon area. The most direct way to find whether black hole microstates are
smooth supergravity solutions is to consider the three-charge black hole, 
which has non-trivial event horizons.   
For these black holes we know \cite{sv} that the statistical
ensemble of the microstates of the boundary theory can exactly account
for the entropy.  It is therefore fascinating  to see if Mathur's picture 
\cite{mathur2}
can be extended to these solutions. If one could find a smooth, three-charge
geometry with no horizon that is dual to {\em every} microstate of the
boundary theory, then our picture of black holes would change
drastically.  A black hole would be nothing more than a classical
effective description of the statistical ensemble of microstate
geometries, and puzzles like the information paradox would be
easily explained.

A few steps towards extending the ``one geometry per microstate''
picture to the three charge case have been made by analyzing these
configurations via the Born-Infeld action of the component branes
\cite{bk}, by perturbing or doing spectral flow on two charge
solutions \cite{mathur2,MathurSV,lunin}, or by constructing families of three
charge solutions with enhanced symmetry \cite{hairs,black}.  So far,
all these steps indicate that the ``one geometry per microstate''
picture of black hole entropy passes quite a few rather non-trivial
consistency checks. However, in order to prove this conjecture, one
needs to find and classify {\em all} supergravity solutions that have
the same supersymmetries and charges as the three charge black hole,
and to map these solutions to microstates of the boundary theory.

As one can see from the physics of three-charge supertubes, the most
generic three charge solutions have three dipole charges that do not
affect their supersymmetry.  Unlike the two-charge supertube solutions
\cite{mathur,supertube,emparan,LuninIZ}, these solutions cannot be found by
usual solution generating techniques, because the three charges can be
generically dissolved into fluxes (as in the Klebanov-Strassler
solution \cite{ks}).  Put differently, the three-charge solutions seem
to be intrinsically non-linear: the Maxwell fields interact non-linearly
with one another because such multi-component solutions cannot avoid the
non-linearities of the supergravity action. 

To find the general three-charge solutions one therefore needs to use more
powerful methods.  The first simplification is to work in a ``duality
frame'' that treats all three charges on the same footing. 
That is, we will work with three sets of
M2-branes that intersect only along the time axis.  One can then
obtain an exact solution of the physically more interesting D1-D5-p
system by compactification and T-dualities.  The other main ingredient
in our search for these solutions is to use directly the fact that
they preserve the same supersymmetries as the three sets of M2
branes that give rise to their asymptotic charges. This is essential if the
solutions are to represent microstates of the three charge  black hole. 
We then make heavy
use of the requirement that mutually BPS brane probes should feel no
force to highly restrict the metric and Maxwell fields of our
backgrounds.  Having done this we use the explicit form of their
Killing spinors to find linear, first-order differential relations
between the metric and the Maxwell fields. As in other
involved supergravity solutions \cite{usc}, knowing the Killing
spinors greatly reduces the complexity of the problem.  We immediately
find that a combination of the Maxwell fields and vielbeins simply has
to be self-dual in the $\IR^4$ transverse to the branes.  We then use
the equations of motion of the four-form to find a set of modified
harmonic equations satisfied by the metric coefficients.  

The resulting system of equations is linear, provided one 
solves it in the right order; at each step the equations are sourced by
quadratic combinations of solutions to previously solved equations.
Note that this makes these solutions amenable to  ``entropy counting:'' 
Since the equations are linear, one can hope to easily superpose and 
combine solutions to create and then count microstates. If the equations 
were non-linear then one could not easily do this, and the non-linearities 
could greatly constrain the solution space. 
Thus the conceptual simplicity of the two-charge solution
persists with three charges. Each step involves solving linear equations 
with known sources,
and the result then completely solves the M-theory equations of motion
while preserving the four supersymmetries of the three underlying
sets of M2 branes. 

The resulting set of equations remind one of the equations satisfied
by all supersymmetric backgrounds of minimal five-dimensional
supergravity \cite{min}, and reduce to these equations when all the
charges and all the dipole charges are respectively equal\footnote{After 
this paper was submitted to the arxiv, we have been informed 
by H.S. Reall that these equations have also been obtained in work that classifies
solutions of five-dimensional $U(1)^3$ invariant supergravity 
\cite{U13}.}.

As we will see, our system of equations allows one to construct huge
classes of solutions, and indicate that one can find a very large
number of BPS black rings. Basically, one is completely free to choose
the dipole charges to have a ring profile of any shape and orientation
in $\IR^4$, and then distribute fundamental (M2-brane) charges with
arbitrary densities along the ring.  Having done this one can find the
corresponding solutions by a linear process.

We illustrate this explicitly for the $U(1) \times U(1)$ invariant black ring. 
The solution is given by specifying three charges, $Q_j$,
three dipole moments, $q_j$, and the radius.  The asymptotic charges are 
the $Q_j$, and the two angular momenta (\ref{ringj}). Given the 
charges and angular  momenta there are thus two free parameters, one of which 
determines the area  of the horizon. Moreover, for each set of asymptotic 
charges there exists a one-parameter family of rings with zero horizon area, 
which are thus three-charge supertubes.  These supertube solutions are regular 
and the Ricci scalar is bounded by $(q_1 q_2 q_3)^{-1/3}$.  Hence, if one 
chooses a solution with sufficiently large dipole charges then the curvature 
can be kept small enough to remain well within the validity of the supergravity 
approximation.
Our solutions  generalize the recent equal charge BPS black ring 
solution \cite{black}, and reduce in near ring limit
to the flat ring metric obtained in \cite{hairs}. 
They also open up the possibility of extending the recent interesting work
on non-BPS black rings \cite{blackring} to solutions without $U(1) \times U(1)$ invariance.

While we have analyzed the black ring in some detail, we stress that
our equations admit solutions with arbitrary profiles.  These solutions
will generically possess non-trivial horizons.  However,
as one approaches the profile that sources the dipoles, the solution will 
look just like the ring solution, and we expect there will be a 
free parameter that will enable us to set the horizon area to zero, 
and recover a completely regular solution with no horizon, but with arbitrary
profile. 

These latter solutions have all the right properties to be the
microstate geometries that give rise to the three charge black hole:
they have the same asymptotic charges, they preserve the same
supersymmetries, they have small curvature everywhere, and have no
close time-like curves or event horizons. One can moreover map 
these geometries to chiral null models \cite{black,6d}, and show they
are exact string backgrounds \cite{null}. Of course, as with all BPS
configurations, there is subtlety in distinguishing simple
superpositions of independent BPS states from the true bound states.
One way to achieve this is to match each bulk solution with a
microstate of the boundary theory and then verify that the latter
really represents a bound state.  However, both the two charge story
and the fact that certain supertubes have a consistent Born-Infeld
description hint quite strongly that three charge supertubes --
solutions with only one ring of dipole charges, all the M2 brane
charges sourced on the ring, and zero entropy -- will be bound states.

If supertube geometries are dual to microstates of the boundary
D1-D5-p CFT, then how are we to interpret the black rings? If the
picture of black-hole hair advanced in \cite{mathur2} is true, then
black rings describe ensembles of these supertube microstates.  The
fact that such a large number of black rings exists strongly supports
this interpretation.  Indeed, one may try to estimate the entropy of
all black rings with fixed charges and angular momenta, and see if
this overcounts the boundary entropy.  If this is true -- and the very
large families of black rings we find suggest it is -- it implies that
a given microstate can be contained into a very large number of black
rings.  The fact that black rings describe sets of nearby microstate
geometries is also a very natural explanation of some of the ring
solutions that can apparently change from black ring to supertube and
back as one moves along the ring.

Aside from the possible duality between the geometries we construct
and microstates of the D1-D5-p system, our solutions are also
interesting in and of themselves, for several reasons.  First, it is
quite amazing that there exists such a large number of BPS solutions
with the same asymptotic charges. Second, some of the explicit
solutions we find describe black rings in ten dimensions. By varying
the black-ring dipole charges we generate a large number of circular
black ring solutions with the same charges. The most general black
ring solution with a given set of asymptotic charges is obtained by
specifying {\em seven} arbitrary functions with a few global
constraints. Hence, these solutions copiously violate black hole
uniqueness.  Third, there are quite a few D-brane
probe calculations that indicate the possibility of creating
geometries with closed time-like curves\footnote{One could do this,
for example, by charging up a three-charge supertube with two dipole
charges \cite{bk,hairs}. Other interesting recent work on 
closed time-like curves in rotating three charge backgrounds has appeared in
\cite{ctc}.}.  Since our solutions should capture {\em
all} BPS three-charge configurations, one can imagine trying to write
down a series of BPS solutions that interpolate between a solution
that has no closed time-like curves and one that has them, to try to
violate chronology protection.  It would be very interesting to see if
this can be done, and, if not, to discover the obstructions.

In section \ref{Equations} we find the equations that give all
supergravity solutions that have the same supersymmetries and charges
as the three-charge black hole, and then in section \ref{Finding} we
give a systematic way to construct generic three-charge solutions, by
solving several inhomogeneous {\em linear} differential equations.
Section \ref{Examples} contains explicit solutions that describe
supertubes and black rings with three charges, three dipole charges,
and $U(1) \times U(1) $ symmetry, as well as solutions containing {\em
both} a BMPV black hole \cite{bmpv} and a black ring.
These solutions
generalize the three equal-charge BPS black ring solutions constructed in
 \cite{black} and reduce in the near ring regime to the solutions in
\cite{hairs}. 
Finally, in Section
\ref{Conclusions} we offer some concluding thoughts and describe
a number of interesting problems that might be addressed using the results
of this paper. Our supergravity conventions
are given in Appendix A, and the BPS brane probe arguments used to constrain 
the metric and the Maxwell fields are given in Appendix B.

{\bf Note Added:} A day after this paper appeared on the arxiv, two more papers appeared
that overlap with our Section \ref{Examples}. In \cite{work}, the solution 
describing a circular BPS black ring with three charges and three dipole charge was found 
and analyzed. In \cite{two2} solutions describing concentric  BPS black rings 
with arbitrary charges and dipole charges in a BMPV black hole background 
were found, generalizing earlier equal-charge multi-BPS-black-ring solutions \cite{two}.

\section{The Equations}
\label{Equations}

In order to find the most general solution, it is best to work in a
duality frame in which the symmetry between the three charges and the
three dipole charges is manifest. We will therefore work in
M-theory with three sets of M2-brane charges, which we take
to be parallel to the $123$, $145$, and $167$ planes respectively\footnote{This
duality frame was first used to describe the three charge black hole in \cite{arc}.}
This is easily dualized to the D1-D5-p system:  One first compactifies
on the 7-direction, and takes the T-dual along the 4 and 5 directions.
This produces a D0-D4-F1 system in IIA supergravity, with the F1 string lying
in the $16$ plane.  Finally, taking the T-dual in the 6-direction produces 
the requisite D1-D5-p system.  M-theory solutions can then be
mapped exactly onto  D1-D5-p solutions by following the Buscher rules.
Note that the directions (8,9,10,11) define an $\IR^4$ that is  inert 
under these dualities and represents the spatial part of the metric in 
which the supertube or black hole is localized. 

From both the Born-Infeld description of three charge supertubes 
\cite{bk,aspects} and from 
the physics of other three-charge geometries \cite{MathurSV,lunin,hairs,black}
we can see that three-charge solutions can also have brane dipole moments.  For
our M2-brane configurations these dipole charges are magnetic and 
correspond to M5 branes in the $12367y$, $12345y$, and
$14567y$, directions, where $y$ denotes a direction in the
$\IR^4$ transverse to all the M2 branes.  Indeed, we will consider
M5 branes that wrap an arbitrary closed curve, $\vec y(\phi)$, in this
$\IR^4$.

Our purpose is to find the most general solution that preserves the same
supersymmetries  as the three-charge black hole.  We can greatly
constrain the form of the metric and the Maxwell fields by 
using the crucial fact that any brane probe that is mutually BPS with the 
three charges of our solutions should feel no force.
These mutually BPS probe branes are: \\ 
$\bullet$ M2 branes
of any orientation in the $\IR^4$ transverse to the tube,
parameterized by $89\, 10\,11$, \\ 
$\bullet$  M2
branes parallel to the charges of the solution, in the $23$, $45$ and $67$ 
directions, \\ 
$\bullet $ M5 branes
which have one direction in the $23$ plane, one in the $45$ plane, one
in the $67$ plane, and the other two directions in the $89~10~11$
hyperplane.

We discuss these probes and analyze the consequences of the 
zero-force condition in  Appendix B of this paper.  Here we simply
note that the most general metric and Maxwell potential Ansatz compatible 
with zero force on the BPS brane probes is:
\bea 
e^1 &=&
e^{-2{A_1} - 2{A_2} - 2{A_3}} \left( {dx}^{1} +
\vec k \cdot d\vec y  \right), \label{e1}\\ 
e^2 &=&
e^{-2{A_1} + {A_2} + {A_3}} {dx}^{2},\\ 
e^3 &=&
e^{-2{A_1} + {A_2} + {A_3}} {dx}^{3},\\ 
e^4 &=&
e^{{A_1} - 2{A_2} + {A_3}}{dx}^{4}, \\ 
e^5 &=&
e^{{A_1} - 2{A_2} + {A_3}}{dx}^{5}, \\ 
e^6 &=&
e^{{A_1} + {A_2} - 2{A_3}}{dx}^{6}, \\ 
e^7 &=&
e^{{A_1} + {A_2} - 2{A_3}}{dx}^{7}, \\ 
e^{7+i} &=&
e^{{A_1} + {A_2} +{A_3}}{dy^i}, \label{e11} \\
C^{(3)} &= & - e^1 \we e^2 \we e^3 - e^1 \we e^4 \we e^5 - e^1 \we e^6 \we e^7 + \\ 
  &+& 2 \,(\vec a_{(1)} \cdot d \vec y)  \we dx^2 \we dx^3 + 2 \,(\vec a_{(2)} 
\cdot  d \vec y )  \we dx^4 \we dx^5 +  2 \,(\vec a_{(3)} \cdot d \vec y)  
\we dx^6 \we dx^7 
\label{cpar} 
\eea
where $\vec y$ parameterizes the $\IR^4$ transverse to the branes, and
$A_1, A_2, A_3, \vec k, \vec a_{(1)}, \vec a_{(2)}$ and $\vec a_{(3)}$ are
functions of the four coordinates $ y^i$. Note that because $e^1$
has some components in the $\IR^4$, the product of vielbeins in $C^{(3)}$ 
implicitly contains components along the transverse $\IR^4$.
We have chosen to write  $C^{(3)}$ in this form because it makes 
significant simplifications to the supersymmetry
variations. The harmonic functions, $Z_i$, that are commonly used to 
write brane metrics are:
\be
Z_i ~\equiv~ e^{6 A_i}\,,
\ee
but these functions are not going to be precisely harmonic here -- they
will have distributed sources.

Note that the functions $Z_i$ implicitly appear in the components
of $C^{(3)}$ parallel to the branes, and thus determine the electric, or
``fundamental'' components of the Maxwell three-form sourced by the monopolar
M2 brane distributions.  The Maxwell one-forms, $\vec a_{(j)}$, on
the $\IR^4$ are normal four-dimensional Maxwell fields on $\IR^4$, but
in the eleven-dimensional theory they govern the magnetic $F^{(4)}$ sourced 
by the M5 brane
distribution of the solution.  So, with a certain abuse of terminology,
we will refer to these as the magnetic, or dipolar Maxwell fields, even though
they are generally electric and magnetic in the four-dimensional
sense.

The key to generating our solutions is that we will construct the
configurations that preserve the same Killing spinors as the three-charge
black hole.  The space of supersymmetries is four-dimensional, and
is defined by the canonical projectors for the M2 branes
(see Appendix A for our conventions):
\be
(\oneone + \Gamma^{123}) \epsilon = (\oneone + \Gamma^{145}) 
\epsilon =(\oneone + \Gamma^{167})  \epsilon =0 \,.
\label{canonprojs}
\ee

Given a supersymmetry, $\epsilon$, in M-theory, the vector
\be
K^\mu ~\equiv~ \bar \epsilon \Gamma^\mu \epsilon 
\label{Kvec}
\ee
must always be a Killing vector \cite{g}. Moreover, it must
be non-zero, and parallel to the time direction. To see the latter,
observe that if $\Delta \equiv \Gamma^{1AB}$ for $AB$ equal to
$23$, $45$, or $67$, then $\Delta$ is hermitian, commutes with
$\Gamma^1$ and satisfies ,  $\Delta^2 =  \oneone$ with 
$\Delta \epsilon = - \epsilon$.  One therefore has 
$ \bar \epsilon \Delta = - \bar\epsilon$, and so
\be
K^a ~=~ \bar \epsilon \Delta \, \Gamma^a \,\Delta \, \epsilon 
~=~  \pm K^a
\ee
where $a$ is a frame index.  To get the last equality one commutes 
$\Delta$ through $\Gamma^a$, and uses $\Delta^2 = \oneone$. One
gets the negative sign if $\Delta$ anti-commutes with $\Gamma^a$,
and so $K^a$ can only be non-zero if and only if it commutes with
all choices of $\Delta$, which only happens for $a =1$.  Moreover,
$K^1 = \epsilon^\dagger \epsilon >0$.  The form of this
Killing vector will be important later, but here we note
that it can be used to normalize the Killing spinor because
$K^1$ (in space-time indices) must be constant, and hence 
the four Killing spinors are
\be
\epsilon_i = e^{-A_1 - {A_2} -  {A_3}}\, \eta_i
\ee
where the $\eta_i$ form a basis of {\it constant spinors} satisfying 
the projection conditions (\ref{canonprojs}).

Since $\Gamma^{123456789\,10\,11} ~=~ \oneone $, it follows that the
Killing spinors must automatically satisfy:
\be
(\oneone - \Gamma^{89\,10\,11}) \epsilon =0 \,.
\label{halfflat}
\ee
In other words, the ``generalized holonomy'' on the transverse
$\IR^4$ must be ``half-flat,'' that is, it must be restricted
to one of the $SU(2)$ factors of $SO(4) \equiv SU(2) \times
SU(2)$.  As we will see, this will lead to a collection
of ``self-duality'' constraints on the background fields
restricted to this $\IR^4$.

The gravitino variation is given in (\ref{gravvar}), and
it is most convenient to consider every variation in terms
of the flat, or frame indices.   There are
three essentially distinct types of variations:  Those
parallel to the spatial sections of the M2 branes, those in the transverse 
$\IR^4$ and the variation in the time direction.  

Because of our choice to split $C^{(3)}$ as in (\ref{cpar}), the
$d\vec k$ terms cancel out of the variations parallel to
the branes.  These variations then give rise to two types of
equations:  The first has already been incorporated into our
Ansatz via the zero-force conditions, and relates the electric 
components of $C^{(3)}$ to the metric functions, $A_j$.  The second
set of equations is a self-duality condition on the ``magnetic''
Maxwell fields. Define the (two-form) field strengths on $\IR^4$ by
\be
G_j ~\equiv~ d (a_{(j)})\,,\quad  j =1,2,3,
\ee
Then the supergravity variations parallel to the spatial 
parts of the M2 branes require that all the differences, 
$Z_i G_i - Z_j G_j$ be self-dual in $\IR^4$.  The variation
along the time direction similarly collapses to the condition
that $\sum_i Z_i G_i$ be self-dual, and so we find that the 
individual field strengths must be self-dual:
\be
G_i ~=~ * G_i~,
\label{formg}
\ee
where $*$ is the {\it flat} Hodge dual on the transverse $\IR^4$. 

The last class of supersymmetry variations relates the vector
field that governs the rotation, $\vec k$, to the Maxwell field, $G_i$:
\be
 d k + * d k  =2 \left( e^{6 A_1}\, G_1  ~+~ e^{6 A_2} \, G_2 ~+~ 
e^{6 A_3} \, G_3  \right) \,.
\label{formk}
\ee

Equations (\ref{formg},\ref{formk}) are all that one needs in order to
satisfy the vanishing of the supersymmetry variations. However, this 
is not sufficient to solve the equations of motions. The simplest to 
solve are the Maxwell equations, and a number of them collapse to trivialities
because of the self-duality of the $G_j$.  The only non-trivial Maxwell
equations arise when two of the $G_j$'s contribute to the $F \wedge F$ term,
and the complete equation requires the functions $Z_i = e^{6 A_i}$ to satisfy 
the amazingly simple equations:
\bea
d * d Z_1 &=& 4 G_2 \wedge G_3~, \nonumber \\
d * d Z_2 &=& 4 G_1 \wedge G_3~, \label{har} \\
d * d Z_3 &=& 4 G_1 \wedge G_2~. \nonumber 
\eea

Imposing these equations solves all of the Maxwell equations.  Fortunately
we do not have to check the Einstein equations because integrability
guarantees that they are automatically satisfied.  This was shown
in \cite{g} to be true whenever the Killing vector, (\ref{Kvec}), 
is time-like, and this is why we were careful to examine the non-zero
components of $K^a$ above, and verify that it is indeed time-like.

Hence, the equations that give {\em all} solutions that 
preserve the same supersymmetry as the three charge black hole are
\bea
G_i & =&  * G_i~ , \label{g1}\\
 d * d Z_i &=& 2 |\epsilon^{ijk}|  G_j \wedge G_k~, \label{g2}\\
d k + * d k  &=&  2 G_1 Z_1 +2 G_2 Z_2 + 2 G_3 Z_3~,
\label{g3}
\eea
where $ |\epsilon^{ijk}|$ is the absolute value of the totally 
antisymmetric tensor.

When all the $Z$'s and $G$'s are equal, these equations reduce to the
equations that yield all supersymmetric solutions of five-dimensional
minimal supergravity \cite{min}.  One can also check that if one
imposes $SO(3)$ symmetry on the $\IR^4$, but allows different charges, 
then the foregoing equations reduce to the those in \cite{hairs}.

While we have taken the transverse space to be $\IR^4$, one can easily
generalize our result by replacing this by any ``half-flat'' 
four-manifold\footnote{The main reason for using a flat base space is the fact that we are
seeking solutions that preserve the same supersymmetries as the three-charge 
black hole in flat space. Moreover, in some regime of 
parameters one expects our solutions to become three charge supertubes, which can also
be described as a D-brane configurations in flat space \cite{bk}. }
That is, we can take the transverse vielbeins to be:
\be
e^{7+i} = e^{{A_1} + {A_2} +{A_3}}{\bar e^i}\,, j=1,\dots,4 \\
\ee
where $\bar e^i$ is the vielbein for any hyper-K\"ahler metric.
Since the curvature tensor on such a space is self-dual, all spinors
satisfying (\ref{halfflat}) will be holonomy singlets and so the 
hyper-K\"ahler connection terms will drop out of gravitino variations 
that involve such spinors.  Thus such a geometry will also lead to a solution 
that preserves four supersymmetries.  While such a replacement makes
no difference to the form of the equations,  non-trivial hyper-K\"ahler 
manifolds have harmonic forms that can provide new components to the
Maxwell fields that make up our solutions.
One can also conjecture a further generalization, where neither the
field strength nor the connection are self-dual, but the combination
of them that appears in the supersymmetry variations is. 


\section{Finding Solutions }
\label{Finding}

While the equations (\ref{g1},\ref{g2},\ref{g3}) exhibit non-linear
relationships between the underlying fields, the actual system that
one has to solve is actually linear at each step, provided that one
solves it in the order presented in (\ref{g1},\ref{g2},\ref{g3}).
As with all the linear systems that underlie brane solutions, there
are choices about where to put the sources, and how to arrange those
sources\footnote{For non-trivial hyper-K\"ahler transverse manifolds
there are also additional choices of harmonic two-forms.}.  At each
step there will thus be the freedom to insert arbitrary source
distributions, and the choices must be guided by the physics that we
wish to describe.  We would, however note that if we want to describe
a bound state then it seems reasonable to assume that all of the
branes must be sourced along the same profile.  We will also see that
near the ring the leading metric coefficients are determined, via
(\ref{g2}), by the components of the three-form flux. Thus the geometry can
sometimes respond to smooth out what could have been singularities at
the sources.

\noindent {$\bullet$ \bf Step 1}

The first equation to solve is the homogeneous one, (\ref{g1}), which
implies that the $G_i$ satisfy the Euclidean, vacuum Maxwell
equations.  Conversely, given a solution, $F$, to the Maxwell
equations one can take $G_i = F + *F$.  Therefore the solution is
determined by the choice of the source distribution and the boundary
conditions at infinity.  (It is also surprising that the equations
that govern the $G_i$ are completely independent of the metric
coefficients $A_i$!)  For the solutions that we seek, the $G_i$ are
sourced by the dipole M5 branes, and so the $G_i$ must fall off
suitably fast at infinity, and the location and shape of these branes
thus completely determines the solution to (\ref{g1}).  Since there
are three independent $G_i$'s, there are three independent source
distributions to be chosen.

The field $G_1$ corresponds to the M5 brane that is parallel to
the $4567$ directions, and occupies a curve in the $\IR^4$.  For
a generalized black ring or supertube, we take this curve to be
a simple closed curve. 
Such a configuration is dipolar and has no net M5 brane charge.
Let $\vec \mu_j (\phi) \in \IR^4$, for $0 \le \phi \le 2\pi$,  define 
the closed curves that source each of the fields, $G_j$ for $j=1,2,3$.
The fields $G_j$ can then be obtained from usual Green function, by first
calculating the vector potentials, $\vec b_{(j)}$, via the line integrals:
\be
\vec b_{(j)}(\vec y) ~\equiv~ \int_0^{2 \pi}\, {\sigma_j(\phi) \, 
{\vec \mu_j}{}' (\phi) \over  | \vec y - \vec \mu_j (\phi)|^2 }   \,
d \phi\,, \label{csource}
\ee
and then setting 
\be
G_j ~=~ (1+*) \,(d \, (\vec b_{(j)}\cdot d \vec y) \,)\,.
\ee
Each of the functions, $\sigma_j(\phi)$, represents the $j^{\rm th}$
M5 brane charge density along the $j^{\rm th}$ curve. Since the M5
branes are spatially extended along these curves one should take each
of these functions to be a constant proportional to the number of M5
branes stacked on each curve.

We should note that at this step we are doing very much the same thing
that was done for the two-charge, one-dipole charge metrics of
\cite{mathur,LuninIZ,emparan}, where the shape of the dipole branes
fully determines the Maxwell fields of these branes.  The only
difference here is that we can, in principle, have different shapes
for the three dipole charges.

For the ``round'' black ring with three equal charges \cite{black}, or
for the more general black rings and supertubes discussed in the next
section, the three $G_i$ are sourced by the same circular profile
lying in a plane in $\IR^4$.  However, we stress that in general, the
singularity profiles of the $G_i$ can be taken to be arbitrary curves,
which by charge conservation must be closed.  Therefore, the most
general singularity profiles will be arbitrarily deformed rings.

At first sight it may seem strange that a rotating brane can have an
arbitrary profile and still be a stable BPS state, however this is
what emerges from our analysis.  Probably the best way to picture it
is that the M5 branes are rotating around their curves at the speed of
light, and are thus rotationally stabilized.  However, at the same
time a traveling wave is running around the M5 brane ring in the
opposite direction, also at the speed of light, so that in the
stationary frame of our metric, this traveling wave is stationary, and
thus gives rise to a general closed curve as a profile.

\noindent {$\bullet$ \bf Step 2}

Having found the $G_i$, one can use equations (\ref{g2}) to find the
metric functions $Z_i$. Given a distribution of branes, the term on
the right hand side of equation (\ref{g2}) contains both the explicit
$G \we G$ term, as well as an implicit ``fundamental'' brane density,
$\rho_i(\vec x)$.  In general, the brane densities, $\rho_i(\vec x)$,
need not be sourced in the same place as the dipole charges that
source the $G_j$.

Since the operator on the left-hand-side of (\ref{g2}) is
the Laplacian on $\IR^4$, it is trivial to invert, and one gets:
\be
Z_i(\vec y) = c_i + \int{ \rho_i(\vec z) + 2 |\epsilon^{ijk}|  
* (G_j \wedge G_k)( \vec z ) \over (\vec y- \vec z)^2 }\, d^4 z \,,
\label{Zsoln}
\ee
where $c_i$ are arbitrary constants that can be set to 1 if one wants 
to obtain asymptotically flat solutions.  Indeed, the boundary conditions
are fixed by specifying the correct asymptotic geometry that contains net 
localized  M2 brane charges.

It is quite remarkable that the combination of the fields sourced by 
two dipoles produces the same effect on the equations as an overall brane
density. A very similar phenomenon happens in the Klebanov-Strassler
solution \cite{ks}, in which the RR and NS-NS three forms transverse to
the D3 branes produce a net D3 brane density dissolved in the fluxes.  
Hence, one can treat the right hand side of (\ref{g2}) as an effective 
brane density coming from branes dissolved into fluxes.

We should note that in the  two-charge, one-dipole charge solution 
\cite{mathur,LuninIZ,emparan} there was no brane density dissolved in 
fluxes, because there was only one dipole charge, and the right hand side of 
(\ref{g2}) was identically zero.

\noindent {$\bullet$ \bf Step 3}

In order to find the rotation parameters, $\vec k$, one must again 
solve a linear inhomogeneous equation (\ref{g3}).

In fact, since this equation only gives the self-dual part of $d k$, 
one can add to $d k$ an arbitrary anti-self-dual form and still 
obtain solutions.  Thus one can, in principle, choose yet another
source distribution, compute its Maxwell field, take the 
anti-self-dual combination of this field, then find the corresponding
vector potential and add that to $\vec k$.  This will also solve
(\ref{g3}).  However, there is important physics in $\vec k$ because
it describes the angular momentum of the solution, and
inserting angular momentum without a corresponding matter
density will generically lead to closed time-like curves.  
Conversely, if one has a matter density, then the freedom to
add a homogeneous solutions of the equation for $\vec k$ can represent
the freedom to spin this matter, and so such homogeneous
solutions can be very important. Once again, though, there will be a
danger of closed time-like curves if $\vec k$ becomes large, or singular.
We will see an example of all this later.  

Therefore, while  adding an  arbitrary anti-self-dual form to $d k$ 
might be possible in the mathematics,  it can be limited or excluded by 
the physics of causality.   

If one acts on (\ref{g3}) with an exterior derivative, one obtains the 
Maxwell's equation for $\vec k$:
\be
* d * d k = * 2 [(dZ_1) \we  G_1  + (d Z_2) \we  G_2  +  (d Z_3) \we G_3 ] 
~\equiv~ J.
\label{J}
\ee
The current $J$ is trivially conserved, and so one can find a solution
to (\ref{J}) by standard methods of electromagnetism. Indeed, one
should note that the change of variables $t \to t + \psi(y)$ induces a
gauge transformation, $\vec k \to
\vec k + \vec \nabla \psi$.  Thus one can work in Lorentz gauge ($*d*k =0$) 
and use the Green function of the Laplacian to define a vector field:
\be
\vec \ell (\vec y) ~\equiv~ \int_{\IR^4}{  \vec J  \over 
| \vec y - \vec z|^2 } d^4 z \,.
\ee 

This vector field is almost $\vec k$.  Consider the two-form 
\be
H ~\equiv~ (1+*) d \ell -  2\, Z_1 \, G_1 - 2\, Z_2\, G_2 - 2\, Z_3\, G_3\,,
\ee
which, by construction, satisfies
\be
d H = d * H = 0 \,,
\ee
even at the sources.  Since on $\IR^4$ the topology is trivial, 
$H$ can always be written as  $ d B$, and hence the most general solution 
to equation (\ref{g3}) is
\be
k = \ell + {B \over 2} + k_{ASD}\,,
\ee 
where $k_{ASD}$ is the undetermined anti self dual part of $k$:
\be
(1+*) d k_{ASD} = 0\,.
\ee

In practice, $H$ is usually going to be zero. Since it is a harmonic
two-form (even at the sources), it is completely determined by the
boundary conditions and the topology.  Since the $Z_i G_i$ fall off
rapidly at infinity, non-trivial $H$'s can't come from
non-normalizable modes, and so $H$ must be zero in $\IR^4$.  It is
amusing to note that there {\it is} a danger for non-trivial topology:
For (\ref{g3}) to be solvable, $\sum_j Z_j G_j$ must be orthogonal to
all harmonic two-forms.

Hence, finding a solution is a three step process, involving the
solving of three sets of {\em linear} equations. The first step is to
find the self-dual forms $G_i$ from the shapes of the dipole
branes. The second is to use this solution and the brane densities to
find the harmonic functions $Z_i$.  The third is to use $Z_i$ and
$G_i$ to find the rotation parameters, $\vec k$.

Of course, actually solving these three sets of equations for generic
dipole profiles is quite non-trivial, and involves non-trivial
integrals.  However, the whole process is linear, and hence presents 
no obstructions. We can also see that near the dipole profile the solution 
becomes that of the three-charge flat supertube, or black tube,
\cite{hairs}. Since that solution has a small Ricci scalar and is free
of singularities, one can infer  that the most general solutions 
will  be non-singular provided that  the three dipole profiles are identical. 
Moreover, when a certain combination of the local charge densities, 
$\rho_j(\vec y)$, and dipole charges is exactly equal to zero, these 
solutions become regular, three-charge supertubes of arbitrary shape.  
When this combination 
is greater than zero, the solutions become black rings of arbitrary shape.
When it is less than zero, the solutions have closed time-like curves, and 
are not physical. 

In the next section we will use the procedure outlined in this section
to construct asymptotically flat  BPS black ring and supertube solutions
with $U(1)\times U(1) $ invariance and arbitrary charges and dipole 
charges. We also construct a solution describing the black ring in 
the background of a rotating three-charge black hole.


\section{Examples}
\label{Examples}

Using our general linear system it is easy to find the most general
black ring solution with three charges, three dipole charges and $U(1)
\times U(1)$ symmetry.  This black ring was conjectured to exist in
\cite{bk} using Mathur's picture of black holes. The near ring limit
of its metric was obtained in \cite{hairs}, and the full ring solution
with equal charges and equal dipole charges was obtained in
\cite{black}.  Other interesting work on black rings has 
appeared in \cite{blackring}.  Our general ring solution can be
further generalized to a solution containing both a black ring and a
BMPV black hole, and so we present both results to illustrate
the power of the method, and to show how the obviously-desirable
superposition principle operates in this system. Since the
non-linearities appear only in source terms for geometric quantities,
we can find the exact non-linear consequences of how two BPS objects
affect one another, and in particular, modify their respective
horizons and entropies.  A different superposition principle for $U(1)
\times U(1)$ invariant geometries with three equal charges has been
used in \cite{two} to obtain superpositions of black tubes and black
holes. It would be interesting to see if that construction extends to
the more generic case considered here.

\subsection{Black rings with arbitrary charges and dipole charges}

If one assumes  $U(1) \times U(1)$ symmetry, then it is natural to pass
to two sets of polar coordinates, $(z,\theta_1)$ and $(r,\theta_2)$ in which:
\be
d\vec y \cdot d \vec y ~=~ (dz^2 ~+~ z^2 \, d \theta_1^2) ~+~
(dr^2 ~+~ r^2 \, d \theta_2^2)  \,.
\ee
We will locate the ring at $r=0$ and $z=R$.
The Maxwell fields, $\vec a_{(j)}$, in (\ref{cpar}) must respect the
$U(1)$ symmetries, and so we take:
\be
\vec a_{(j)}\cdot d \vec y ~=~   c_j(z,r)\,  
d \theta_1 + d_j(z,r) \, d\theta_2\,, 
\ee
where $c_j$ and $d_j$ are arbitrary functions.  In principle, there 
could be  $z$-components and $r$-components to the $\vec a_{(j)}$, 
leading to a
field strength term proportional to $dr \wedge dz$.  However,
self-duality and the Bianchi identities mean that there would
have to be a term exactly of the form $const. \, d\theta_1 \wedge d\theta_2$,
and this would lead to singular geometry at infinity. 

While we could continue to work on these coordinates, it is better
to work in a form of bipolar coordinates that makes the ring appear
simpler, and most particularly leads to the simplest possible expressions
for the self-dual $G_j$.  We want the $G_j$ to be simple because they 
generate the sources for the other equations in our linear system.   

The best coordinate system to use is the one in \cite{black}, in which the 
flat $\IR^4$ metric has the form:
\be
ds^2_{\IR^4}= {R^2 \over (x-y)^2}\left( {dy^2 \over y^2 -1} 
+ (y^2-1) d \theta_1^2 
+{dx^2 \over 1-x^2} + (1-x^2) d \theta_2^2  \right).
\ee
and the self-dual\footnote{Our 
orientation is $\epsilon^{yx\theta_1\theta_2} = 1$.} 
field strengths are
\be
G_i =  q_i (d x \we d \theta_2 - d y \we d \theta_1) \,.
\ee
The actual change of variables is:
\bea
x &=& -{ z^2+r^2 - R^2 \over  \sqrt{((z-R)^2 + r^2)( (z+R)^2 + r^2 )}} \,, \\
y &=& -{ z^2+r^2 + R^2 \over  \sqrt{((z-R)^2 + r^2)( (z+R)^2 + r^2 )}} \,,
\eea
and if we express the $G_i$ in terms of the $z$ and $r$ coordinates,
we find exactly the Maxwell forms sourced by a circular, 
uniform dipole profile 
in (\ref{csource}).  The $(x,y)$ coordinates lie in the ranges $ -1 \le x
\le 1$, $-\infty < y \le -1$, and the ring is located at
$y = -\infty$.

Since the $G_j$ are so simple, it is quite an easy exercise to find the form 
of the $Z_i$ that satisfy (\ref{g2}):
\bea
Z_1 &=& 1 + {Q_1 \over R} (x-y) - {4 q_2 q_3 \over R^2}(x^2 - y^2) ~,\\
Z_2 &=& 1 + {Q_2\over R} (x-y) - {4 q_1 q_3 \over R^2}(x^2 - y^2) ~,\\
Z_3 &=& 1 + { Q_3\over R} (x-y) - {4 q_1 q_2 \over R^2}(x^2 - y^2) ~.
\eea
The terms proportional to $q_j q_k$ are sourced by $G_j \wedge G_k$, while
the terms proportional to $Q_i$ correspond to choosing a 
uniform ``fundamental''
charge distribution, $\rho_i$, in (\ref{Zsoln}).  Indeed, one should note
that the functional dependence of these terms is:
\be
x-y = {2 R^2 \over \sqrt{((z-R)^2 + r^2)( (z+R)^2 + r^2 )}}
\ee
which shows that the $Q_i$ terms are precisely
sourced by a circular ring  of radius $R$ with constant charge densities,
and that the $Q_i$ are proportional to the charge densities.

The self-dual two-form that appears on the right hand side of (\ref{g3}) is
\be
2 \sum_{i=1}^3 \,Z_i\, G_i   ~=~ [A + B(x-y) + C(x^2 -y^2)]  \, 
(d x \we d \theta_2 - d y \we d \theta_1) \,,
\ee
where
\be
A \equiv 2 (q_1+q_2+q_3)~,~~~ B \equiv {2\over R} (Q_1 q_1 + Q_2 q_2 + 
Q_3 q_3)~,~~~ C \equiv -  {24 q_1 q_2 q_3 \over R^2}.
\ee
Hence, the components of the one-form, $k$, satisfy the equations:
\bea
(y^2-1){\p_y k_{2} } &+& (1-x^2){\p_x k_{1}  } = 0 \,, \label{eq1}\\
{\p_x k_{2} } - {\p_y k_{1}  } &=&  A + B(x-y) + C(x^2 -y^2) \,. \label{eq2}
\eea

If we now define $p_1$ and $p_2$ by:
\be
k_{1} ~=~ p_1 ~+~ ( \alpha -1 ) A y ~+~ \frac{1}{2} B y^2 + 
\frac{1}{3} C y^3\,, \qquad
k_{2} ~=~  p_2~+~ \alpha A x + \frac{1}{2} B x^2 + \frac{1}{3} C x^3 \,,
\label{pdefns}
\ee
for some freely choosable parameter $\alpha$, then one has
\be
(1- y^2){\p_y  p_2 } - (1-x^2){\p_x p_1 } ~=~ 0 \,, \qquad
{\p_x  p_2 } - {\p_y p_1} ~=~ 0 \,.\label{redkeqns}
\ee
One can then differentiate and eliminate either $ p_1$ or $p_2$.
Indeed, eliminating $ p_2$ yields a simple partial differential
equation for $p_1$:
\be
(1- y^2){\p_y^2 p_1} - \p_x((1-x^2){\p_x p_1 }) ~=~ 0 \,.
\label{ktwopde}
\ee
This is trivially separable, and one is led to the ODE's:
\be
(1- x^2) \,\frac{d^2}{dx^2} f(x) - 2x \,\frac{d}{dx} f(x)  + \lambda\,
f(x) ~=~ 0 \,, \qquad (1- y^2) \,\frac{d^2}{dy^2} g(y) + 
\lambda\, g(y) ~=~ 0\,,\label{ODEs}
\ee
where $\lambda$ is a separation constant.   The first of these equations
is the Legendre equation, and if one differentiates the second equation
with respect to $y$ then one sees that $g'(y)$ must also satisfy the Legendre
equation. 

The precise form of Legendre function depends upon the boundary conditions.
Recall that $-1 < x <1$ and $-\infty < y < -1$, and:
\bea
x = -1 \ &\Leftrightarrow&  \ r=0\,, \ z > R   \,; \qquad
x = +1 \ \ \Leftrightarrow \  \ r=0\,, \ z < R \\
y = - \infty \ &\Leftrightarrow&  \ r=0 \,; \ z=R  \,,  \qquad
y = - 1 \ \ \Leftrightarrow \ \ z=0\,.
\eea
We do not want any singularities inside or outside the ring and
so $f(x)$ must be regular at $\pm 1$.  This means that 
$\lambda=n(n+1)$ for some integer, $n \ge 0$, and $f(x) =P_n(x)$,
where $P_n(x)$ is the corresponding Legendre polynomial.
Given that $\lambda=n(n+1)$, the function $g'(y)$ must be 
a combination of $P_n(y)$ and the second Legendre function, 
$Q_n(y)$.  However the latter has a logarithmic singularity
at $y=-1$, and hence we have $g'(y) = P_n(y)$.  

Define $\widehat P_n(y) = \int P_n(y) dy$, where the arbitrary 
constant of integration is fixed by requiring that $\widehat P_n(y)$
satisfy the second equation of (\ref{ODEs}). Then the
most general regular solution to  (\ref{ktwopde}) is of the form:
\be
p_1~=~ \sum_{m=0}^\infty \, a_n\, P_n(x)\,  \widehat P_n(y) \,,
\qquad p_2 ~=~ \sum_{m=0}^\infty \, a_n\,  \widehat P_n(x) \, P_n(y)
\,,\label{gensol}
\ee
where $a_n$ are arbitrary constants.   We can now fix all of these 
constants by examining the asymptotics and requiring that there be no 
time-like curves near the ring.

The ring is located at $r=0$, $z=R$, or at $y= -\infty$, and the
surface of the ring can be described by $(x,\theta_1,\theta_2)$. The
$(r,z)$ coordinates are singular in this region, indeed setting $z = R
+ \epsilon$ and $r = |\epsilon \tan \psi|$, one finds that, to zeroth
order in $\epsilon$, $x = - \cos \psi$.  Thus one of the coordinates
on the surface of the ring involves the ratio $r/(z-R)$.  We will thus
use the $(x,y,\theta_1,\theta_2)$ coordinates and since $ -1 \le x \le
1$ globally, we will set $x= -\cos \psi$.

Along the ring, the metric reduces to the three-metric:
\be
ds_3^2 ~=~ {V^2 \, R^2 \over (x-y)^2}\, \Big[\, (y^2-1)\, d \theta_1^2 ~+~
d \psi^2 ~+~ \sin^2 \psi \, d \theta_2^2 \,\Big] ~-~ V^{-4}\,
(k_1\,d \theta_1 + k_2\,d \theta_2)^2
\,,\label{ringmet}
\ee
where
\be
V ~\equiv~ (Z_1 Z_2 Z_3)^{1/6} \,.
\ee
The obvious danger is that if the second term dominates then it is
possible to have closed time-like curves, and this danger is most
acute when  $y \to -\infty$.  In this limit one has $V^2 \sim
(\frac{C^2}{9\,R^2})^{1/6} y^2$, and so $p_1$  can, at most, be
cubic in $y$ and $p_2$ can, at most, be quadratic in $y$.  
Since $\widehat P_2(y)$ is actually a cubic, this means that 
one must take $a_n =0 , n\ge 3$ in (\ref{gensol}).  It turns out
that one must also take $a_2 =0$.  This is because the $y^2 d \theta_1^2$
terms are proportional to $C^2 - (C + \gamma a_2 P_2(x))^2$ for some
constant, $\gamma$.  Near the tube,  $x$ can vary from -1 to 1, 
depending on the approach angle, $\psi$, and because $P_2(x)$ changes
sign in this range, there will always be a range of $x$ values
in which closed time-like curves occur unless we take
$a_2 =0$. Thus the $y^2 d \theta_1^2$ terms must cancel exactly. 

Next, there are terms of the form $y d \theta_1^2$, and these have a
coefficient proportional to $x \left( {C \over 3} - {a_1 \over 2}
\right)$.  Again, since $-1 < x < 1$ near the ring, there will always
be a region in which this term is dominant and negative, giving rise
to closed time-like curves, unless we take ${a_1 \over 2} = {C \over 3}
$.  The constant $a_0$ can be absorbed into the parameter $\alpha$
introduced in (\ref{pdefns}). It is also possible to add arbitrary
constants, $b_j$, to each of the $p_j$, but these can be fixed by
requiring that the $k_j$ vanish at infinity.

There is one last subtlety in the issue of closed time-like curves.
Above we have collected the leading terms in the metric in the limit
$y \to - \infty$.  However, some of these leading terms vanish at
$x=\pm 1$, and so we must examine sub-leading terms, and even though
they are vanishing as $y \to - \infty$, they can still give rise to
closed time-like curves.  Indeed, one finds such problematic terms at
$x=1$, and they are proportional to $- \alpha^2$. This means that
one must also set $\alpha \equiv 0$. The term proportional to 
$\alpha$ an anti-self-dual contribution to $\vec k$ sourced on the ring, 
and corresponds to spinning the ring in the transverse directions. Our 
solution indicates that this is not possible, despite the presence of a
nontrivial matter density on the ring\footnote{One can also show
that the flat black ring is dual to the four-charge four-dimensional 
BPS black hole. Hence,
the fact that $\alpha$ must be zero is not surprising, since a
non-zero $\alpha$ corresponds to giving  this  black hole angular momentum.}.

The end result is: 
\bea
k_{1} &= &   (y^2-1)\,\left({C \over 3} \,(x+y) ~+~
{B \over 2} \right) ~-~  A \,(y+1) \,, \nonumber \\
k_{2} &= &    (x^2-1)\,\left({C \over 3}\, (x+y)~+~
{B \over 2} \right) .
\label{ksol2}
\eea

In terms of usual $\IR^4$ coordinates the solution for the black ring 
can be written as: 
\bea
Z_i & = & 1 + {2 R Q_i \over \Sigma} 
+ {8 |\epsilon^{ijk}| q_i q_j (r^2+z^2) \over \Sigma^2 } \,,\\
k_1 &=& {4 R^2 z^2 \over \Sigma^2} 
\left( {Q_1 q_1 + Q_2 q_2 + Q_3 q_3 \over R} 
+{16 q_1 q_2 q_3 (r^2+z^2)\over R^2 \Sigma}  \right)  + 
{8 R^2 (q_1+q_2+q_3) z^2 \over \Sigma (\Sigma + r^2 + z^2 + R^2) }\,, \\
k_2 &=& -{4 R^2 r^2 \over \Sigma^2} 
\left( {Q_1 q_1 + Q_2 q_2 + Q_3 q_3 \over R} 
+ {16 q_1 q_2 q_3 (r^2+z^2)\over R^2 \Sigma}  \right)\,, \\
C^{(3)}  &=& - e^1 \we e^2 \we e^3 - e^1 \we e^4 \we e^5 
- e^1 \we e^6 \we e^7 + 
\nonumber \\ 
  &+& \left( { z^2+r^2 + R^2 \over  \sqrt{((z-R)^2 + r^2)( (z+R)^2 + r^2 )}} 
\, d \theta_1 -{ z^2+r^2 - R^2 \over  \sqrt{((z-R)^2 + r^2)( (z+R)^2 + r^2 )}} 
\, d \theta_2 \right)  \nonumber \\
 &  & \we\, ( 2 q_1  dx^2 \we dx^3 + 2 q_2 dx^4 \we dx^5 
+  2 q_3 dx^6 \we dx^7) \,,
\eea
where $\Sigma$ is the inverse of the harmonic function sourced by the ring:
\be
\Sigma \equiv {\sqrt{((z-R)^2 + r^2)( (z+R)^2 + r^2 )}} \,.
\ee

With these expressions for $k_j$, one can
easily check that the three-metric (\ref{ringmet})  has the following
limit as $y \to - \infty$:
\be
ds_3^2 ~=~ \bigg({C^2 \over 9\, R^2 }\bigg)^{1 /3} \, \bigg[\,
\bigg({9\, R^2 \over C^2 }\bigg)\,M\, d \theta_1^2 ~+~ R^2\,( 
d \psi^2 ~+~\sin^2 \psi \, (d\theta_1 + d\theta_2)^2) \,\bigg]  \,,
\label{tubemet}
\ee
where 
\be
M ~\equiv~  (2 q_1 q_2  Q_1 Q_2 + 2 q_1 q_3  
Q_1 Q_3 + 2 q_2 q_3   Q_2 Q_3 - q_1^2 Q_1^2 - q_2^2 Q_2^2 -
q_3^2 Q_3^2)  + \frac{1}{3} \, A \, C \,R^2\,.
\label{Ddefn}
\ee
We therefore see that the ring does indeed have the geometry
of $S^2 \times S^1$, and that the horizon has a volume of 
$8 \pi^2 M \, R^2$ and a cross-sectional area of
$4 \pi ({1\over 9} C^2 R^4)^{1/3}$. This near-ring metric reproduces exactly
the metric of the flat ring found in \cite{hairs} after identifying 
$q_i = d_i$. 

The supertube solutions are those with $M=0$, and hence have zero horizon
area.  As $y \to -\infty$, the warp factors in front of the spatial part
of the M2 branes go to a finite limit of the form $(q_i q_j/q_k^2)^{1/3}$,
while the remaining five-dimensional part of the space-time becomes
$AdS_3 \times S^2$, albeit with a null orbifold in the $AdS_3$. 
The full eleven-metric is thus  regular for $M=0$.
It is also important to note that at least the Ricci scalar,
and presumably all the other curvature invariants, are of 
order $(q_1 q_2 q_3)^{-{1\over 3}}$, 
and so even relatively modest dipole charges lead to a solution
whose curvatures can be kept away from the string scale.  Therefore
the supergravity approximation can be trusted as a description
of these backgrounds, and thus, if Mathur \cite{mathur2} is correct, as
a description of black-hole microstates.

One can also express the entropy and the angular momenta 
of the black ring in terms of the quantized ring charges, $\bar N_i$,
and dipole charges $n_i$, which are related to $Q_i$ and $q_i$ via:
\be
Q_i = {\bar N_i l_p^6 \over 2 L^4 R}\,,\qquad q_i={n_i l_p^3 \over 4 L^2}\,,
\ee
where $L$ is the length of the two-tori. The asymptotic charges, 
$N_i$, of the solution are the sum of the charges 
on the black ring $\bar N_i$, and the charges dissolved in fluxes:
\be
N_1 = \bar N_1 + n_2 n_3 ~,~~~ N_2 = \bar N_2 + n_1 n_3 ~,~~~ 
N_3 = \bar N_3 + n_1 n_2  \,.
\label{N}
\ee

Similarly, the angular momenta have both a contribution from the ring, 
and a contribution from the fluxes:
\bea
J_1 &=& J^T ~+~ 
\left(n_1 n_2 n_3 ~+~ \sum_{i=1}^3\, {n_i \bar N_i \over 2} \right) ~=~ 
J^T ~+~ {1\over 2}
\left( \sum_{i=1}^3\, {n_i N_i} ~-~ n_1 n_2 n_3 \right)\\
J_2 &=& -\left( n_1 n_2 n_3 ~+~ \sum_{i=1}^3\,  {n_i \bar N_i \over 2} \right)
~=~ - {1\over 2}
\left( \sum_{i=1}^3\, {n_i N_i} ~-~ n_1 n_2 n_3 \right)
\label{ringj}
\eea
where $J^T$ is the angular momentum carried by the ring. Even if  both $J^T$ 
and $n_i$ are quantized, supersymmetry requires them to be related:
\be
J^T  = {R^2 L^4 \over l_p^6}\,(n_1+n_2+n_3)\,.
\label{j}
\ee
This relation determines the radius of the ring.

The entropy is simply
\be
S = {2 \pi A \over \kappa_{11}^2} = \pi \sqrt{\cal M} 
\ee
where 
\bea
{\cal M}& =& 
2 n_1 n_2 \bar N_1 \bar N_2 +2 n_1 n_3 \bar N_1 \bar N_3  
+2 n_2 n_3 \bar N_2 \bar N_3 -  n_1^2 \bar N_1^2 -  n_2^2 \bar N_2^2 
-  n_3^2 \bar N_3^2  - 4 (J_1 + J_2) n_1 n_2 n_3  \nonumber \\
&=& 2 n_1 n_2  N_1  N_2 +2 n_1 n_3  N_1  N_3  
+2 n_2 n_3  N_2  N_3 -  n_1^2  N_1^2 -  n_2^2  N_2^2 
-  n_3^2  N_3^2  \\
&-&  n_1 n_2 n_3 [4(J_1 + J_2) + 2 (n_1 N_1 + n_2 N_2 + n_3 N_3)  - 3  n_1 n_2 n_3 ]
\nonumber
\eea
We can see that given the three asymptotic charges $N_i$ and the 
two angular momenta, there is still a two parameter family of circular 
black rings with those charges.  (There are seven parameters, $(N_i,n_j,R)$ 
and five asymptotic charges.) Hence, the circular black rings {\it alone}
copiously violate black-hole uniqueness.

\subsection{A black ring in a rotating black hole background}

In order to describe a black ring in the background of a rotating
three-charge black hole, one must add to the harmonic functions $Z_i$
a component coming from a black hole at the origin. If the black hole
has charges $ Y_1, Y_2$ and $Y_3$, the harmonic functions become:

\bea
Z_1 &=& 1 - {Y_1 \over R^2}{x-y \over x+y } + {Q_1 \over R} (x-y) 
- {4 q_2 q_3 \over R^2}(x^2 - y^2) ~,\\
Z_2 &=& 1 - {Y_2 \over R^2}{x-y \over x+y }+ {Q_2\over R} (x-y) 
- {4 q_1 q_3 \over R^2}(x^2 - y^2) ~,\\
Z_3 &=& 1 - {Y_3 \over R^2}{x-y \over x+y }+ { Q_3\over R} (x-y) 
- {4 q_1 q_2 \over R^2}(x^2 - y^2) ~.
\eea
and the equations satisfied by $k$ get an extra term from the new source:
\bea
(y^2-1){\p_y k_{2} } &+& (1-x^2){\p_x k_{1}  } = 0 \label{bhk1}\\
{\p_x k_{2} } - {\p_y k_{1}  } &=&  A + B(x-y) + C(x^2 -y^2) + 
D {x-y \over x+y} 
\label{bhk2}
\eea
where 
\be
D \equiv - { 2 \over R^2} (Y_1 q_1 + Y_2 q_2 + Y_3 q_3)\,.
\label{D}
\ee
The solution is:
\bea
k_{1} &= &   (y^2-1)\,\left({C \over 3} \,(x+y) ~+~
{B \over 2}  + {D \over x+y} + {K \over R^2 (x+y)^2 }\right) 
~-~  \, A \,(y+1)  \,,\nonumber \\
k_{2} &= &    (x^2-1)\,\left({C \over 3}\, (x+y)~+~
{B \over 2} + {D \over x+y} + {K \over R^2 (x+y)^2 } \right)\,. 
\label{bhsol}
\eea
Here we have added an extra homogeneous solution of the equation for
$k$.  That is, the term proportional to $K$ gives an anti-self-dual
contribution to $dk$ and is thus annihilated by  $(1+*)d$.  Physically,
this term represents the angular momentum of the BMPV black hole. 
In principle we could have added this term to the pure black ring 
solution as well. However, there will be closed time-like curves
if one introduces angular momentum without accompanying mass density.
More precisely, this new term will introduce closed time-like curves
unless there is a black hole of charges $Y_1 Y_2 Y_3 > K^2$ at the origin.
The sign of $K$ is undetermined, so the black hole can spin 
in the same, or in the opposite direction to the black ring.

The charges of the black hole are related to actual number of 
branes via
\be
N^{\rm BH}_i = {Y_i L^4 \over l_p^6}\, , 
 ~~~~~~  J^{\rm BMPV} = {K L^6  \over l_p^6}\, .
\ee

By looking at the asymptotics of the solution it is quite easy to 
find the angular momenta:
\bea
J_1 &=& J^T ~+~\left[ n_1 n_2 n_3 ~+~ \sum_{i=1}^3 \left( 
{n_i \bar N_i \over 2} ~+~ n_i N^{\rm BH}_i \right) 
~+~  J^{\rm BMPV}\right]\, \\
& = &  J^T ~+~\left[ - { n_1 n_2 n_3 \over 2} ~+~ \sum_{i=1}^3 \left( 
{n_i N_i \over 2} ~+~ n_i N^{\rm BH}_i \right) 
~+~  J^{\rm BMPV}\right]\,, \\
J_2 &=&  -\left[ n_1 n_2 n_3 ~+~ \sum_{i=1}^3 \left(
{n_i \bar N_i \over 2} ~+~ n_i N^{\rm BH}_i\right) ~+~ J^{\rm BMPV}\right] \\
& = &  -\left[ - { n_1 n_2 n_3 \over 2} ~+~ \sum_{i=1}^3 \left( 
{n_i N_i \over 2} ~+~ n_i N^{\rm BH}_i \right) 
~+~  J^{\rm BMPV}\right]\,,
\eea
where $J^{\rm BMPV}$ is the angular momentum of the BMPV black hole at
the origin, and $J^T$ is the angular momentum of the black ring
(\ref{j}).  The other angular momentum terms come from the fluxes, and
include an interaction term between the black hole and the ring.

As expected, the solution asymptotes to the BMPV black hole solution
near the black hole, and to the circular black ring solution
(\ref{tubemet}) near the ring. The entropy of the black hole is
unchanged; however, the non-linear interaction between the black hole
and black ring (\ref{D}) modifies the entropy of the latter:
\be
S = {2 \pi A \over \kappa_{11}^2} = \pi \sqrt{\cal M} 
\ee
where 
\bea
{\cal M} &=& 
2 n_1 n_2 \bar N_1 \bar N_2 +2 n_1 n_3 \bar N_1 \bar N_3  
+2 n_2 n_3 \bar N_2 \bar N_3 -  n_1^2 \bar N_1^2 -  n_2^2 \bar N_2^2 
-  n_3^2 \bar N_3^2  \nonumber \\
&-&  4 n_1 n_2 n_3 (J_1 + J_2 + n_1 N^{\rm BH}_1 + n_2 N^{\rm BH}_2 
+ n_3 N^{\rm BH}_3 ) \,.
\eea

\subsection{Black rings and supertubes of arbitrary shapes}

Thus, for every set of asymptotic charges and angular
momenta, there is a discretuum of circular black rings and supertubes with
the same asymptotic charges. 

Besides circular black rings our results show that there will be a
very large of number of black rings with the same asymptotic charges,
and no symmetry. As we have argued in section \ref{Finding}, any
closed curve in $\IR^4$ gives a solution, which, near the curve, has
the same metric as the flat black ring. To avoid closed time-like
curves transverse to the tube one needs to avoid adding terms in the
kernel of the $(1+*)d$ operator that become singular at the
tube\footnote{These terms are similar to the terms proportional to
$\alpha$ in our ``round tube'' solution.}. Hence, given the charges,
dipole charges and tube shape there is no freedom in changing $k$.  To
avoid closed time-like curves along the tube one must have a large
enough charge density, so that $M$ is always non-negative.

There is another way to see that the circular black rings constructed
here can be deformed to an arbitrary shape.  It was shown by Horowitz and
Marolf \cite{hm} that one can add a traveling wave along a four-charge,
five-dimensional black string while keeping the metric smooth and
supersymmetric. The flat four-charge black string is dual to our
black ring in the large radius limit, and a traveling wave corresponds to
a ripple on the ring profile. Since the ripple and the black ring move
with the speed of light in opposite directions, the resulting
configuration is a static deformation of the black ring.

One can also check that smoothly varying the charge densities along a
flat tube \cite{hairs} does not cause any problems to the solutions,
as long as the constraint $M \ge 0$ is everywhere satisfied. This is
not surprising, since the leading contribution to the metric near the
tube comes from the dipole moments, and not from the M2 brane
charges. Since the near ring
metric always approaches the flat tube one, we conclude that black
ring solutions where the charge densities vary along the ring are good
solutions.

Hence, we expect a generic three-charge, three-dipole charge black
ring to be determined by seven functions -- four embedding functions
and three charge densities, satisfying the constraint $M>0$ everywhere
on the ring.  In addition to these seven functions, we also have three
discrete parameters - the dipole charges.  Thus there are a huge
number of black rings, and as we mentioned in the introduction. it
would be interesting to see if they overcount the entropy of the
boundary D1-D5-p system.

To go from black rings to supertubes with three charges, three dipole 
charges and an arbitrary shape we only need to change the requirement 
$M>0$ to $M=0$ everywhere along the tube. This implies that a generic 
supertube metric is determined by six functions. Since the supertubes have 
no entropy and low curvature everywhere, their number might be large enough
to account for all the microstates of the D1-D5-p system.

It also appears to be possible to obtain solutions in which $M=0$ in
some sectors of the ring, and $M>0$ in others. These solutions would
describe black hole beads on a supertube. It would be interesting to
fully analyze these exotic configurations, and see if one can freely
change the horizon topology by simply moving charges around the
ring. If those solutions are free of pathologies, they would provide
an interesting laboratory for studying horizon topology change in a
controlled BPS setting.

\section{ Conclusions and Future Directions}
\label{Conclusions}

We have analyzed all eleven-dimensional supergravity solutions that
preserve the same supersymmetries as the three-charge black hole.
Besides the three asymptotic charges, the interior of the solutions
may contain up to three dipole moments, associated to other types of
branes.  We also found that the general solution is completely
determined by a simple linear system of equations that is equivalent
to solving a set of charge distribution problems in four-dimensional
electromagnetism.  The most generic, regular solution involves seven
arbitrary functions and is given by first specifying a closed curve,
or ``generic ring'' of arbitrary shape in $\IR^4$.  The four arbitrary
functions describing this ring determine the shape of the dipole
branes.  The remaining three arbitrary functions represent the
densities of each of the three fundamental brane-charges and how they
are spread around the ring.

Near the ring, the metric approaches the flat-ring metric of
\cite{hairs}. Hence, depending on how large the charge densities are,
the solutions can be either black rings of arbitrary shape, or three
charge supertubes of arbitrary shape.  In section (\ref{Examples}) we
have constructed several solutions with $U(1) \times U(1)$ symmetry,
including a circular black ring/three-charge supertube with arbitrary
charges and dipole charges, and the black ring in a rotating BMPV
black hole background.  We have also argued that, since a solution is
obtained by solving linear equations, any arbitrary ring profile in
$\IR^4$ with three arbitrary charge density functions gives a
solution.  Moreover, near the ring the metric is free of pathologies
provided the charge densities are larger than a lower bound; therefore
these solutions are regular at least down to and across the horizon.
These black rings can have the same charges and angular momenta as the
BPS three charge black hole, and thus copiously violate black hole
uniqueness\footnote{While this might have been expected of black rings
\cite{blackring}, the magnitude of the violation is definitely beyond 
expectation.}.

If the three charge density functions satisfy a constraint, the
solutions have zero horizon area, and thus have no entropy. They are
also completely regular. Moreover the scalar curvature in the core is
bounded by the inverse of the cubic root of the product of the dipole
charges. Thus the supergravity approximation is valid, and these
solutions are good string theory backgrounds. It is also possible
to map at least some of these geometries to chiral null models
\cite{black,6d}, and thus show they receive no string corrections
\cite{null}.

While the equations are linear, the solutions to one set of equations
feed non-linearly into the sources of subsequent linear equations, and
so the fundamental charge distributions depend non-linearly upon the
dipolar distributions, while the gravitational background depends
non-linearly upon all the charge distributions.  When more than one
dipole charge is present, the Maxwell fields sourced by the dipole
branes give rise to dissolved brane charges, much like in the
Klebanov-Strassler solution \cite{ks}. Hence, the harmonic functions
appearing in the metric receive a contribution both from the actual
charge densities on the ring (which may vary along the ring), and from
charges ``dissolved'' in the dipole fields. Similarly, the angular
momentum of the solution contains both a direct contribution from the
ring and a contribution from the Maxwell fields sourced by the dipole
branes.

Our systematic construction also allows one to find solutions
describing arbitrary superpositions of black rings, black holes,
supertubes, and more exotic objects. Of course, the more
complicated the shape of these objects, the more difficult it is to
solve the inhomogeneous differential equations that give the
solutions. On the other hand, the underlying system of equations is
simple, and relatively well understood, and so there are many
interesting, general calculations that might be done.  For example, it
would be very interesting to do some ``maximum entropy'' calculations
to find the most probable configurations with given asymptotic
charges.  One might also examine what happens when tubes cross, or
touch -- this might describe a transition from a bound state to a set
of states.

One of the most important things we may hope to do is to map all the
zero-entropy, regular solutions to the microstates of the D1-D5-p
system. If this is successful, then, as argued by Mathur \cite{mathur2}, 
our picture
of black holes would change at the most fundamental level: black holes
would be understood as statistical ensembles of regular microstate
geometries.  Alternatively, one could try to count the regular
solutions (in a similar way to the counting of two-charge supertubes
\cite{bd,ohta}), and see if there are enough of them to account for
the black hole entropy. Of course, to really prove this conjecture,
one must first determine whether our ring configurations represent
true bound states.  Geometry {\it does} seem to indicate a natural
candidate for bound states, namely, the three-dipole charge supertubes
with fundamental charges localized on the tube.  This seems natural,
not only by analogy with the two-charge system but also because many
of these tubes have a consistent Born-Infeld
description \cite{bk,aspects}. Nevertheless, it would be good to 
see this from the
perspective of the dual CFT.

However, before doing any counting or mapping to the CFT, the fact
remains that we have such a huge number of black rings -- seven free
functions worth of them. This may already be a very strong hint that
this picture of black holes is valid.  In this context, we also think
it significant that the underlying equations are linear.  This is
because black hole entropy can be described by combinations of branes,
and to make larger black holes one simply needs to combine more
branes.  This picture seems to implicitly need some form of
superposition principle, and if there were non-linearities then one
might get non-trivial restrictions on the phase space of one group of
branes imposed by another group of branes.  Counting would be a
nightmare! At the very least, the linearity of the underlying system
will make the statistics much easier to analyze.

If black holes are ensembles of geometries, then so are black rings; a
black ring should be the statistical ensemble of nearby supertube
geometries with the same charges. However, if black rings truly
describe {\em ensembles} of microstate geometries, then it should be
possible for a certain geometry to be part of more then one
ensemble. If this is correct, then the entropy of the black rings with
a fixed asymptotic charges should be larger than the corresponding
D1-D5-p system entropy.  It would be really interesting to see if this
is indeed the case\footnote{When all charges are equal the entropy of
each black ring is of order $N$, so at first glance, the family of
black rings classified by seven free functions seems to definitely
overcount the entropy.}. This picture also makes the more exotic black
hole beads configurations easier to understand: they are simply
ensembles of microstate geometries that are fixed in some regions, and
allowed to vary in others.  The ``beads'' are where the ensemble
averages are being taken, while the supertubes sections are where the
microstate is fixed.

One can also use our solutions to construct all four-dimensional
three-charge solutions that are mutually BPS with the four-dimensional
black hole.  The three-charge limit of this black hole has no entropy;
hence, the entropy of the dual CFT cannot be described by the
black-hole geometry.  As in the D1-D5 system \cite{mathur}, it should be
possible to find the microstate geometries that account for this
entropy.

In addition to trying to establish Mathur's conjecture, there are quite
a number of interesting related questions.  One should try to understand 
the complicated entropy formula of circular black rings using the 
D1-D5-p system.  What is the corresponding boundary description? What
features of these microstates correspond to the dipole charges of the
black ring? Given a black ring with an arbitrary shape, how can one find
the corresponding boundary sector?  

Then there are questions related to more classical gravitational
physics.  Most of the solutions that we have found have angular
momentum, and are very close to having closed time-like curves. This
is particularly evident in the supertube with two dipole
charges. Consider the the near-ring metric parameter, $M$, in
(\ref{Ddefn}) for $q_3 = 0$: One finds $M = - (Q_1 q_1 -Q_2 q_2)^2$.
Thus one would have closed time-like curves unless $(Q_1 q_1 -Q_2 q_2)
\equiv 0$, in which case the solution is a supertube.  However, the
Born-Infeld analysis \cite{bk} indicates that there is no force needed
to bring a fundamental charge near a two-dipole supertube, but if one
drops such a charge into the two-dipole supertube then the resulting
solution would have closed time-like curves.  It would be in very
interesting to construct the full geometry that describing this
process, and to see if there is any obstruction to this apparent
violation of chronology protection.

It is also very intriguing to analyze the more exotic configurations
that have a non-zero horizon size only in several sectors of the
ring. Such solutions would describe black hole beads on a supertube,
or pinched horizons, like a string of sausages. Our solutions indicate
that one can go from a horizon of nonzero size to one of zero size by
simply moving charges around the ring, and might provide a controlled
BPS setting for studying horizon topology change.

Another series of solutions one might try to construct would describe
the dropping of a black ring into a black hole. Again the Born-Infeld
analysis \cite{bk} suggests that this is possible, and that the end
result would be a BPS black hole with unequal angular
momenta. Nevertheless, it has been argued
\cite{harvey-6} that such black holes do  not exist.  

We therefore believe that the results and ideas presented here will
find a lot of interesting applications.  We believe that we have found
the most general ring solution with the same supersymmetries as the
three-charge black hole, and thus we indeed have the ``One Ring to
Rule Them All.''  Probably the most intriguing of all the questions
outlined here is whether they all represent bound states, and whether
the black hole can be thought of as a bound-state ensemble average of
the regular supertube microstates.  That is, does our One Ring really
``in the darkness bind them?''

\subsection*{Acknowledgments}

I.B.~ would like to thank Per Kraus for very stimulating discussions
and insights.  N.W.~ would like to thank the Aspen Center for Physics
for its hospitality and for providing a very stimulating atmosphere
while this work was completed.  This work was supported in part by
funds provided by the DOE under grants DE-FG03-84ER-40168 and
DE-FG03-91ER-40662, and by the NSF under grants PHY00-99590 and
PHY01-40151.  \footnote{Any opinions, findings and conclusions
expressed in this material are those of the authors and do not
necessarily reflect the views of the Lord of the Rings, of the
U.S. Department of Energy or of the National Science Foundation.}

\section*{Appendix A:  Supergravity Conventions}

The metric is  ``mostly plus,'' and we take the gamma-matrices to be
\bea
\Gamma_1 &~=~&  -i \, \Sigma_2 \otimes 
\gamma_9 \,, \quad
\Gamma_2~=~  \Sigma_1 \otimes \gamma_9 \,, \quad 
\Gamma_3~=~   \Sigma_3 \otimes \gamma_9 \,,  \\
\Gamma_{j+3} &~=~&  {\oneone}_{2 \times 2} \otimes \gamma_j \,, 
\quad j=1,\dots,8 \,, 
\eea
where the $\Sigma_a$ are the Pauli spin matrices, $\oneone_{2 \times 2} $ is
the identity matrix, and  the $\gamma_j$ are real, symmetric $SO(8)$
gamma matrices.  As a result, the $\Gamma_j$ are all real, with 
$\Gamma_1$ skew-symmetric and $\Gamma_j$ symmetric for $j>2$.  One also
has:
\be
\Gamma^{1 \cdots\cdots 11} ~=~ \oneone \,,
\ee
where $\oneone$  denotes the $32 \times 32$ identity matrix.

The gravitino variation is 
\be
\delta \psi_\mu ~\equiv~ \nabla_\mu \, \epsilon ~+~ \coeff{1}{288}\,
\Big({\Gamma_\mu}^{\nu \rho \lambda \sigma} ~-~ 8\, \delta_\mu^\nu  \, 
\Gamma^{\rho \lambda \sigma} \Big)\, F_{\nu \rho \lambda \sigma}
\label{gravvar}  \,.
\ee
With these conventions, sign choices and normalizations,
the equations of motion are:
\bea
R_{\mu \nu} ~+~ R \, g_{\mu \nu}  &~=~&  \coeff{1}{12}\, 
F_{\mu \rho \lambda \sigma}\, F_\nu{}^{ \rho \lambda \sigma}\,,
\label{EinEqn}\\ 
\nabla_\mu F^{\mu \nu \rho  \sigma} &~=~& -  \coeff{1}{1152} \, 
\varepsilon^{\nu \rho  \sigma \lambda_1\lambda_2
\lambda_3 \lambda_4 \tau_1\tau_2  \tau_3 \tau_4} \,F_{ \lambda_1  \lambda_2
 \lambda_3  \lambda_4} \, F_{ \tau_1\tau_2 \tau_3 \tau_4} \,.
\label{MaxEqnA}
\eea
The Maxwell equation may be written more compactly as:
\be
d*F ~+~ \coeff{1}{2} \, F \wedge F ~=~ 0 \,. \label{MaxEqnB}
\ee

\section*{Appendix B:  Brane probe constraints}

In this Appendix we constrain the form of the general solution
that preserves the same supersymmetries as the three-charge black hole
by using the crucial fact that any brane probe that is mutually BPS
with the black hole should feel no force.  The first observation is
that the bulk metrics must maintain the same isometries as the
boundary. The three M2 branes are wrapped on $T^2 \times T^2 \times
T^2$, and so any metric that mixes the $T^2$'s with other directions
is excluded. Moreover, there cannot be any four-form field strengths
with an odd number of legs along any $T^2$.

The solution has M2 brane charges in the 123, 145 and 167 planes.
There are three types of probe branes that are mutually BPS with these
charges.  First, there are the M2 branes carrying these charges.
Second, we can put an M2 brane along any two directions in the
four-dimensional space transverse to the branes. The projector
associated with these branes commutes with the three projectors which
determine the Killing spinors of our solutions, and the resulting
configuration preserves two supercharges. Third, we can put an M5
brane that has one direction in the $23$ plane, one in the $45$ plane,
one in the $67$ plane, and the other two directions in the $89\, 10\,
11$ hyperplane. To see that this configuration is BPS one can pick
without loss of generality the M5 brane coordinates to be 35789. If we
reduce along 7, T-dualize along 8, S-dualize, and T-dualize twice
along 2 and 5 we obtain a configuration with four D3 branes, in the
358, 248, 256 and 239 directions, which again preserves two
supercharges.

The transverse M2 branes are BPS regardless of which two of the four
transverse direction they span.  There are no Maxwell fields that
couple to these branes, as they would correspond to giving the
solution a fourth charge. Therefore, the determinant of the induced
metric is constant. This implies that the $dx^1 \we dx^i \we dx^j$
component of $e^1 \we e^i \we e^j $ is constant, and therefore all
transverse vielbeins are equal:
\be
e^i_i = {1 \over \sqrt{ e^1_1}} 
\ee 
Of course, there is nothing preventing $e^1$ from having components
along the $\IR^4$.  These components do not enter any brane action,
and will only be determined by the full supersymmetry equations. Hence
the transverse metric will have a flat base.

The fact that one can rotate one of the M5 brane directions in the
three $T^2$'s on which the M2 branes giving the charges are wrapped
implies that the two vielbeins of each $T^2$ are equal. Moreover,
since there is no Maxwell field coupling with this brane, the zero
force condition implies that the product of the vielbeins of the three
tori is one.

We now use the fact that branes along the 23, 45 or 67 directions feel
no force.  This implies that the product of the vielbeins $e_1^1 e_2^2
e_3^3 $ is equal to the Maxwell potential $C_{123}$, and similarly for
the 45 and 67 vielbeins.  This completely establishes the form of the
metric and of the electric Maxwell potentials to be the one in
(\ref{e1}--\ref{e11}).

Determining the magnetic Maxwell fields is more involved. First, the
$T^2 \times T^2 \times T^2$ isometry implies that all four-form field
strengths can have either two legs on the same $T^2$, or no legs on
the $T^2$'s, or all four legs on two different $T^2$'s.  Moreover, one
can argue that forms with no legs on the $T^2$'s make an M2 brane in
the transverse $\IR^4$ feel a non-zero force; also, the supergravity
equations of motion in the three-charge background imply that turning
on a field strength with all four legs along two $T^2$'s induces a
form with no legs on the $T^2$'s, which the brane probe analysis
excludes.  Therefore, the only Maxwell fields compatible with our
supersymmetries have two legs along either of the three two-tori. This
determines the form of the Maxwell field Ansatz (\ref{cpar}).

The above argument has only one gap, coming from the fact that a supergravity 
brane probe analysis cannot be used to find the force between two branes that 
can be dualized to the D0-D8 system. These D-branes are mutually BPS, and yet a 
naive analysis the Born Infeld action of one of the brane in the background of
the other  gives a nonzero force. This is because there exists a RR interaction 
between these branes that is not captured by supergravity (for a more detailed analysis of 
this system see \cite{d0d8}). In our case the branes whose interaction with the probe 
M2 branes is not captured by supergravity would be D6 branes that lift to KK monopoles in 
M-theory. It would be interesting to see if allowing metric factors of the type these 
branes source would allow for more general solutions dual to black hole hair.


\end{document}